\documentstyle[12pt,a4,epsf]{article}

\newcommand{\wt}{\widetilde}
\newcommand{\ol}{\overline}
\def\mod{\mathop{\rm mod}\nolimits}
\def\diag{\mathop{\rm diag}\nolimits}
 
\begin{document}

\baselineskip=18pt plus 0.2pt minus 0.1pt

\begin{titlepage}
\title{\hfill\parbox{4cm}
       {\normalsize YITP-99-24\\{\tt hep-th/9905059}\\May 1999}\\
       \vspace{3.5cm}
       String junctions on backgrounds\\
       with a positively charged orientifold plane
       \vspace{2cm}}
\author{Yosuke Imamura\thanks{\tt imamura@yukawa.kyoto-u.ac.jp}
\\[7pt]
{\it Yukawa Institute for Theoretical Physics,}\\
{\it Kyoto University, Kyoto 606-8502, Japan}
}
\date{}

\maketitle
\thispagestyle{empty}

\vspace{1cm}

\begin{abstract}
\normalsize
By means of the heterotic/type IIB duality,
we study properties of junctions on backgrounds with a positively charged
orientifold seven-plane and D-branes,
which are expected to give seven dimensional $Sp(r)$ gauge theories.
We give a modified BPS condition for the junctions
and show that it reproduces
the adjoint representation of the $Sp(r)$ group.
\end{abstract}


\vfill

\noindent

\end{titlepage}

\section{Introduction}
Recent years the string theories have been bringing us a lot of information
on non-perturbative properties of various gauge theories.
In the context of the string theories,
gauge theories are described as theories on branes.
Among several ways to realize four dimensional field theories
by brane configurations, in this paper,
we shall focus on the construction with D3-brane probes on
7-brane backgrounds.
Particles in a field theory on the D3-branes are represented by
string junctions connecting some of the 7-branes and the probe D3-branes.
Because the flavor symmetry of the field theory is also regarded
as the gauge symmetry on the 7-branes,
it is very important to study how to construct gauge
theories on 7-branes with various gauge symmetries.

It is already known that
we can construct theories with arbitrary simply-laced gauge groups
and analyze their spectra by introducing $[p,q]$
7-branes\cite{fromopen,arbitrary,affine,uncovering,km,flavor,const,Geo}.
The BPS condition for states in the field theories is
represented with the self-intersection numbers of corresponding junctions.
Using this condition, we can determine the spectra of field theories
with ADE flavor symmetries even in the strong coupling region.
However, the junction description of four
dimensional field theories
with non-simply-laced flavor symmetries
in the strong coupling region
has not been established.
Especially, the BPS condition has
not yet been obtained.

On the other hand, in the weak coupling limit,
we can easily construct $Sp(r)$ gauge theories
as gauge theories on a stack of parallel $r$ D-branes and an orientifold seven-plane
with positive R-R charge ($O^+$).
The purpose of this paper is to study junctions on this background
by means of the heterotic/type IIB duality.

One might expect that one can construct $SO(2r+1)$ gauge theory
by using an orientifold seven-plane with negative R-R charge ($O^-$), $r$ D7-branes
and a fractional D7-brane, which is a D7-brane identified with its mirror.
However, because this set of 7-branes has the non-integral monodromy
\begin{equation}
M_{SO(2r+1)}=\left(\begin{array}{cc}
                   -1 & 7/2-r \\
                      & -1
             \end{array}\right),
\end{equation}
this configuration is not allowed as a classical solution
of supergravity.
From the viewpoint of a field theory on $n$ probe D3-branes,
this fact implies that ${\cal N}=2$ $Sp(n)$ gauge theory with
$r+1/2$ hypermultiplets in the fundamental representation has an anomaly.
(Classically, we can consider half-hypermultiplets.
They are hypermultiplets whose scalars and fermions belong to
$({\bf2},{\bf2n})$ and $({\bf1},{\bf2n})$
representations of $SU(2)_R\times Sp(n)_{\rm gauge}$ respectively.)
It is nothing but the Witten's topological anomaly\cite{su2anomaly}.
Therefore we will not consider $SO(2r+1)$ gauge theories on 7-branes.
Furthermore, although 7-brane constructions of
other non-simply-laced groups ($F_2$ and $G_2$)
are also interesting problems, we will not argue them and
we will concentrate our attention to the $Sp(r)$ group.

To obtain $Sp(r)$ gauge theories on 7-branes,
we should lay an $O^+$ and $r$ D7-branes in parallel.
This configuration has the following monodromy.
\begin{equation}
M_{Sp(r)}=\left(\begin{array}{cc}
                   -1 & -4-r \\
                      & -1
             \end{array}\right).
\end{equation}
Because this monodromy is identical to that of a set of $8+r$ D7-branes
and an $O^-$,
one might expect the BPS condition for junctions on
these two backgrounds to be identical.
However, in the latter, an open string going around $O^+$
with its both endpoints on one D7-brane cannot be BPS state,
and it is undesirable because
the string corresponds to the Cartan part of the $Sp(r)$ gauge group.
This implies that we should modify the BPS condition.

In this paper, we analyze the BPS condition by means of
the hetero/IIB duality.
In \cite{JuncHet}, the correspondence between quantum numbers
of heterotic strings on ${\bf T}^2$ and
ones of junctions on ${\bf T}^2/{\bf Z}_2$ is obtained.
Because all the orientifold seven planes of the ${\bf T}^2/{\bf Z}_2$
used in \cite{JuncHet} are $O^-$, we cannot use the result
for our present purpose.
In the case in of ${\bf T}^2/{\bf Z}_2$ containing an $O^+$,
the dual is heterotic string theory
compactified on ${\bf T}^2$ with `twisted' gauge bundle\cite{WithoutVector}.
In Section \ref{sec:Dualities}, we review this duality,
and we explain how the BPS condition for heterotic strings should be modified
in Section \ref{sec:BPS}.
In Section \ref{sec:1to1}, we give the one to one correspondence
between quantum numbers of heterotic strings and ones of junctions
and suggest the modified BPS condition for junctions.
The last section is devoted to conclusions and discussions.

\section{Heterotic/Type IIB duality}\label{sec:Dualities}
In this section we will explain the duality between $SO(32)$\footnote{%
Precisely, the gauge group is not $SO(32)$ but $Spin(32)/{\bf Z}_2$.
In this paper, we will not care such a point about the discrete subgroups.}
heterotic string theory
compactified on ${\bf T}^2$ with non-trivial gauge bundle structure
structure and type IIB theory on ${\bf T}^2/{\bf Z}_2$
with three $O^-$ and one $O^+$ \cite{WithoutVector}.
Let us start from the heterotic string theory
whose $X^7$ and $X^8$ directions are compactified on ${\bf T}^2$
with radii $R_7$ and $R_8$, respectively.
For simplicity,
we assume that the ${\bf T}^2$ is rectangular
and that the NS-NS 2-form field $B_{78}$ vanishes.
Let us consider the maximal subgroup $O(2)\times O(16)$ of $SO(32)$.
In order to realize the nontrivial structure of the gauge bundle
on the torus, we should introduce Wilson lines for both the $O(2)$ and
the $O(16)$ as follows:
\begin{equation}
\mbox{$X^7$ direction} : \sigma_z\otimes\exp(2\pi iA^a_7H_a),\quad
\mbox{$X^8$ direction} : \sigma_x\otimes\exp(2\pi iA^a_8H_a).
\label{sowl}
\end{equation}
The factors $\sigma_z$ and $\sigma_x$ are $O(2)$ Wilson lines,
which are nothing but Pauli matrices for $\bf2$ representation.
These are essential for the non-trivial structure of the gauge bundle.
(In what follows, we represent this torus by ${\bf T}^2_\ast$,
while ${\bf T}^2$ means a torus with trivial structure of the bundle.)
Due to these Wilson lines, the gauge group $SO(32)$ is broken to
$O(16)$.
$H_a$ and $A^a_\mu$ ($a=1,\ldots,8$, $\mu=7,8$)
are the Cartan generators and Wilson lines
of this $O(16)$ respectively.

First, we should carry out the S-duality transformation
to obtain type I theory on the ${\bf T}^2_\ast$.
Then, we should perform T-duality transformation along the $X^7$ direction
to have type IIA theory.
Because the $X^7$ direction is compactified on ${\bf S}^1$
with radius $R_7$ on the type I side, the type IIA theory is compactified on
${\bf S}^1/{\bf Z}_2$ with length $\pi\wt R_7=\pi/R_7$.
The Wilson line $\sigma_z\in O(2)$ along the $X^7$ direction
breaks the gauge group $SO(32)$ to its subgroup $O(16)\times O(16)$
(Here, we assume $A_\mu^a=0$.)
and each $O(16)$ factor lives on the eight D8-branes on top of one of two
orientifold eight-planes ($O8$).
The $X^8$ direction is compactified on ${\bf S}^1$ with radius $R_8$ as it was in the type I theory.
Because of the existence of the non-trivial structure of gauge bundle on
the ${\bf T}^2_\ast$,
the compactification manifold of type IIA theory is not just a direct product
of the ${\bf S}^1/{\bf Z}_2$ and the ${\bf S}^1$.
Non-commutativity between the Wilson lines along the $X^7$
and the $X^8$ direction,
\begin{equation}
\sigma_x\sigma_z\sigma_x=-\sigma_z,
\end{equation}
implies that when we go around the ${\bf S}^1$ along the $X^8$ direction,
we should swap the two $O8$.
As a result, we have a M\"obius strip as a compactification manifold
of type IIA theory.

\begin{figure}[htb]
\begin{displaymath}
\put(70,20){$\pi\wt R_7$}%
\put(-25,75){$2\pi R_8$}%
\put(43,75){$A$}
\put(103,75){$B$}
\put(240,-10){$\pi\wt R_7/2$}%
\put(185,77){$4\pi R_8$}%
\put(255,105){$A$}
\put(255,45){$B$}
\put(73,-30){$(a)$}
\put(253,-30){$(b)$}
\epsfbox{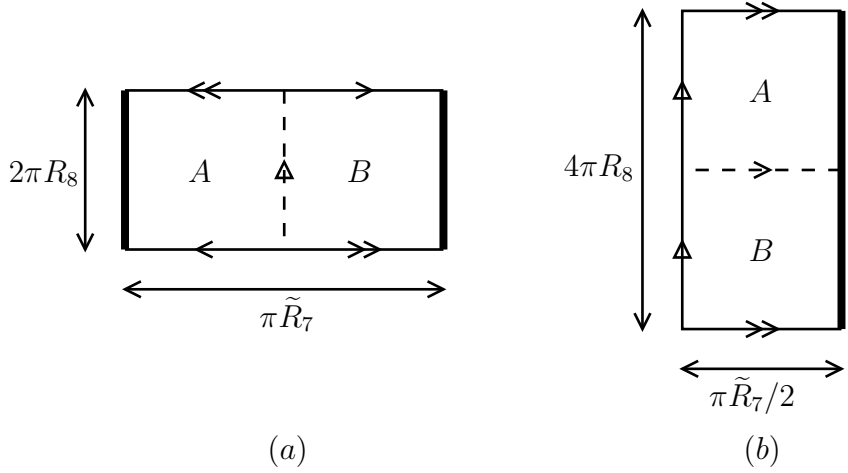}%
\end{displaymath}
\caption{A M\"obius strip (a) can be rearranged to a cylinder
with one end capped with a cross cap (b).
The thick lines represent orientifold eight-planes.}
\label{mobius.eps}
\end{figure}

In order to perform further T-duality transformation along the $X^8$
direction to arrive at the final type IIB
configuration, it is convenient to rearrange the M\"obius strip to a cylinder.
Namely, as is shown in Fig.\ref{mobius.eps},
the M\"obius strip with width $\pi\wt R_7$
and circumference $2\pi R_8$ ((a) in Fig.\ref{mobius.eps})
is equivalent to a cylinder with length $\pi\wt R_7/2$ and circumference $4\pi R_8$
whose one boundary is $O8$ and another is capped with a cross cap ((b) in
Fig.\ref{mobius.eps}).
After this rearrangement, it becomes clear that the $X^8$ direction is compactified
on ${\bf S}^1$ with radius $2R_8$ (not $R_8$).
By performing T-duality transformation along this ${\bf S}^1$,
we obtain ${\bf T}^2/{\bf Z}_2$ with radii $\wt R_7/2$ and $\wt R_8/2=1/(2R_8)$.
(Radii of ${\bf T}^2/{\bf Z}_2$ mean radii of two cycles of the ${\bf T}^2$.)
By this T-duality transformation,
the cross cap is transformed into an $O^-$
and an $O^+$,
and the $O8$ on the boundary
is transformed into two $O^-$.
We will refer this compactification manifold by ${\bf T}^2/{\bf
Z}_2^\ast$,
while orientifold with four $O^-$ by ${\bf T}^2/{\bf Z}_2$.

Using this series of dualities, we can obtain some properties of $O^+$.
For example, let us consider a type I D-string winding once around the $X^7$ direction,
which is S-dual of a heterotic string winding around the same direction.
By the first T-duality along the $X^7$ direction, this D-string is transformed into
a fractional D-particle on $O8$, whose mirror image is itself,
and by the second T-duality along the $X^8$ direction it is
transformed into a type IIB D-string with winding number $1/2$.
The winding number $1/2$ implies that the D-string is stretched
between two orientifold planes along the $X^8$ direction.
In the cylindrical representation of the M\"obius strip ((b) in Fig.\ref{mobius.eps}),
the $O8$, on which the fractional D-particle exists, is only on one side.
Therefore, the type IIB D-string is
also on the same side of the ${\bf T}^2/{\bf Z}_2^\ast$.
This fact suggests that a D-string cannot be attached to $O^+$.
We can obtain the same conclusion by starting from a type I D-string
wrapped around the $X^8$
direction as follows:
Because the gauge group on the D-string is $O(1)={\bf Z}_2$,
the Wilson line on it takes two values $\pm1$.
Through the T-duality along the $X^7$ direction
the D-string is transformed into a D2-brane wrapped over the M\"obius strip
which inherits the Wilson line of the D-string.
If we rearrange the M\"obius strip to the cylindrical representation,
the length of the cycle along the $X^8$ direction become twice.
As a result, the Wilson line is squared and always becomes $+1$
regardless of the original value.
Therefore, the possible $X^8$ coordinate of the dual type IIB D-string
is unique.
This is consistent with the fact that a D-string cannot be attached to $O^+$.
Of cause,
although one D-string cannot be attached to $O^+$,
even number of D-strings can be simultaneously attached
because they can be regarded as D-strings going around $O^+$ and turning back.
Furthermore, the number of fundamental strings attached to an
orientifold plane is, regardless of its charge, always even integer.
As a result, we conclude that both F-string charges and D-string
charges of strings attached to $O^+$ should always be even integers.

\begin{figure}[htb]
\begin{displaymath}
\put(90,-10){$X^7$}
\put(-15,100){$X^8$}
\put(60,60){$(M_0,0)$}
\put(115,80){$(M_1,N_1)$}
\put(75,130){$(M_2,N_2)$}
\put(25,105){$(M_3,N_3)$}
\put(0,0){$O^+$}
\put(0,170){$O^-$}
\put(170,0){$O^-$}
\put(170,170){$O^-$}
\put(55,25){$D7\times8$}
\epsfbox{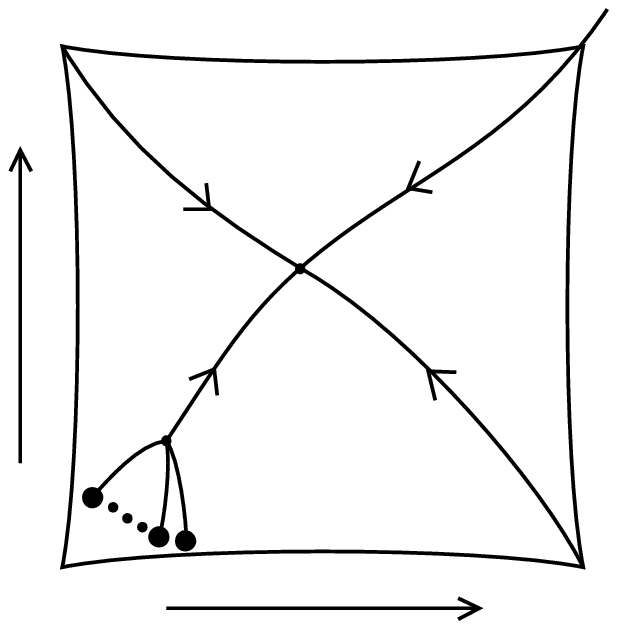}
\end{displaymath}
\caption{The standard configuration.}
\label{standard.eps}
\end{figure}

In the case of junctions on ${\bf T}^2/{\bf Z}_2$ (with four $O^-$),
we can always remove strings attached
to one of four $O^-$ by adding null junctions \cite{JuncHet},
which can be shrinked to a point by a continuous deformation.
However, in that case, we should admit improper junctions, which
contain strings with fractional charge,
in order to remove a string with odd integral charge attached on the $O^-$.
On the other hand,
because the F-string charge and D-string charge of a string attached
on $O^+$ are always even integers as we have explained,
for any junction on ${\bf T}^2/{\bf Z}_2^\ast$,
we can remove a strings attached on $O^+$
without introducing such improper junctions.
Therefore, we can deform any junction on ${\bf T}^2/{\bf Z}_2^\ast$
into `a standard configuration',
in which strings starting from three $O^-$ planes and eight D-branes
on $O^+$ meet together at a point as is expressed in Fig.\ref{standard.eps}.
(If it is necessary, we should move the D7-branes.)
In this configuration, arbitrary junctions are specified by
F-string charges $M_1$, $M_2$, $M_3$ and D-string charges $N_1$,
$N_2$, $N_3$ of strings going
outward from three $O^-$ and the number of strings $Q_a$ attached on $a$-th
D7-brane.
By means of the charge conservation, the following relations hold:
\begin{equation}
M_0+M_1+M_2+M_3=0,\quad
N_1+N_2+N_3=0,
\end{equation}
where $M_0$ is the sum of $Q_1,\ldots,Q_8$.
Therefore, we have $12$ independent charges.

Due to the monodromies of 7-branes,
we should introduce branch cuts
and we cannot define the string charges until
we fix the branch cuts.
Our convention is as follows:
The ${\bf T}^2/{\bf Z}_2^\ast$ is made of two rectangles,
i.e., a rectangle on the front side and that on the back side.
We assume that all the branch cuts run on the back side
and that all strings run on the front side.
Using this definition, we can define the string charges
up to transformation $\tau\rightarrow\tau+1$.

\section{The BPS condition}\label{sec:BPS}
On the worldsheet of heterotic strings,
in the light cone formalism,
there are eight bosonic fields $X_L^i$ and eight fermionic fields $\psi^i$
as the left movers
and $24$ bosonic fields $X_R^i$ as the right movers.
(We assume the left mover is supersymmetric.)
Among the right movers, $16$ bosonic coordinates are compactified on the
self-dual even lattice and are equivalent to $32$ fermionic fields
through the fermionization.
Let us use $\lambda^1,\ldots,\lambda^{32}$ to represent these $32$ fermions.
The $SO(32)$ symmetry among these real fermions gives the spacetime gauge symmetry.

At first, let us consider the heterotic string theory compactified on
${\bf T}^2$ (with four $O^-$)
along the $X^7$ and the $X^8$ directions.
For the argument below, it is convenient to combine these
real fermions into complex fermions as follows:
\begin{equation}
\lambda^{a+}=\lambda^{2a-1}+i\lambda^{2a},\quad
\lambda^{a-}=\lambda^{2a-1}-i\lambda^{2a},\quad
a=1,\ldots,16.
\label{combine2}
\end{equation}
The $\lambda^{a\pm}$ carry a charge $\pm1$ with respect to $a$-th $U(1)$ factor
of Cartan subgroup $U(1)^{16}\subset SO(32)$.
Let ${\bf A}^{(16)}_\mu=(A_\mu^1,\ldots,A_\mu^{16})$ ($\mu=7,8$)
denote the Wilson lines
for the $U(1)^{16}$.
Because the fermions couple to them,
the boundary condition of the fermions
on a string with winding number $n_\mu$ depends on the Wilson line.
\begin{equation}
\lambda^{a\pm}(\sigma+2\pi)=\exp2\pi i(c\pm n_\mu A_\mu^a)\lambda^{a\pm}(\sigma),
\label{lambdabc}
\end{equation}
where $c=0$ for R-sector and $c=1/2$ for NS-sector.
Let $\lambda^{a\pm}_m$ denote the oscillators of $\lambda^{a\pm}$ with frequency $m$,
satisfying the following commutation relations:
\begin{equation}
\{\lambda^{a+}_m,\lambda^{b+}_n\}=
\{\lambda^{a-}_m,\lambda^{b-}_n\}=0,\quad
\{\lambda^{a+}_m,\lambda^{b-}_n\}=\delta_{a,b}\delta_{m+n}.
\end{equation}
According to the boundary condition (\ref{lambdabc}),
the frequency $m$ of $\lambda^{a\pm}_m$ takes values in ${\bf Z}+c\pm n_\mu A^a_\mu$.
The charges $q^a$ associated with the $a$-th $U(1)$
is given by
\begin{equation}
q^a
=\sum_{m\in{\bf Z}+c+n_\mu A^a_\mu}\lambda^{a+}_m\lambda^{a-}_{-m}.
\label{pdef}
\end{equation}
If we use the $\zeta$ function regularization, this is rewritten as
\begin{equation}
{\bf q}^{(16)}={\bf k}^{(16)}-{\bf A}^{(16)}_\mu n_\mu,
\label{qandk}
\end{equation}
where ${\bf q}^{(16)}$ is $(q^1,\ldots,q^{16})$
and ${\bf k}^{(16)}$ is a vector of normal ordered charges taking values on self-dual even lattice
generated by
\begin{equation}
{\bf e}_1-{\bf e}_2,\quad
{\bf e}_2-{\bf e}_3,\quad
\ldots\quad,\quad
{\bf e}_{15}-{\bf e}_{16},\quad
{\bf e}_{15}+{\bf e}_{16},\quad
\frac{1}{2}\sum_{i=1}^{16}{\bf e}_i,
\end{equation}
where ${\bf e}_i$ are vectors such that their set
$\{{\bf e}_1,\ldots,{\bf e}_{16}\}$ is the orthonormal basis.
The operators $L_0$ and $\wt L_0$ are given as follows:
\begin{equation}
L_0=\frac{1}{4}\sum_{\mu=7,8}(p^\mu-w^\mu)^2+N,\quad
\wt L_0=\frac{1}{4}\sum_{\mu=7,8}(p^\mu+w^\mu)^2+\wt N,
\end{equation}
where $p^\mu$ and $w^\mu$ are Kaluza-Klein momenta and winding lengths
along $X^\mu$
quantized by $w_\mu=n_\mu R_\mu$ and $p^\mu=m'_\mu/R_\mu$, respectively, where
\begin{equation}
m_\mu'=m_\mu+{\bf k}^{(16)}\cdot{\bf A}^{(16)}_\mu
      -\frac{1}{2}{\bf A}^{(16)}_\mu\cdot{\bf A}^{(16)}_\nu
      n^\nu,\quad
m_\mu,n_\mu\in{\bf Z}.
\label{mpis}
\end{equation}
The dots in (\ref{mpis}) represent the inner products defined with
metric $g_{ab}={\bf1}_{16}$.
Concerning the right movers,
$\wt N$ is decomposed as
\begin{equation}
\wt N=\wt N_{\rm osc}+\frac{1}{2}{\bf q}^{(16)}\cdot{\bf q}^{(16)}-1,
\end{equation}
where $\wt N_{\rm osc}$ represents the contribution from
excitations which do not change the fermion number ${\bf q}^{(16)}$.
Because $N$ should vanish for BPS states and
$\wt N_{\rm osc}$ is always non-negative,
we obtain the following BPS condition from the level matching condition $\wt L_0=L_0$.
\begin{equation}
2m'_7n_7+2m'_8n_8+{\bf q}^{(16)}\cdot{\bf q}^{(16)}\leq2.
\label{bpscond}
\end{equation}
Using integral indices $m_\mu$, $n_\mu$  and ${\bf k}^{(16)}$, we have
\begin{equation}
2m_7n_7+2m_8n_8+{\bf k}^{(16)}\cdot{\bf k}^{(16)}\leq2.
\label{bps}
\end{equation}
This is the BPS condition for heterotic strings on ${\bf T}^2$.
Because these heterotic strings are related to the junctions
on ${\bf T}^2/{\bf Z}_2$, we can interpret the condition (\ref{bps})
as that for junctions.

When we establish the relations between quantum numbers of
heterotic strings and invariant charges of the dual type IIB junctions,
we should carefully take account of
the depencende of the fundamental string winding numbers
upon the position of D-branes
as is mentioned in \cite{BGL} for type I' case.
If there exists a fundamental string attached on a winding D-string,
the winding number of the fundamental string
depends on the position of the D-string.
To avoid this matter,
as is done in \cite{JuncHet},
let us use the flat background
on which the R-R charge is locally cancelled.
On the background,
the fundamental string winding numbers are independent of
the positions of D-strings and
the normalized Kaluza-Klein momenta $m_\mu'$ and winding numbers $n_\mu$ are
identified with the F-string winding numbers $F_\mu$ and D-string winding numbers $D_\mu$
in the type IIB picture by the following simple relations:
\begin{equation}
F_7=m'_7,\quad
F_8=m'_8,\quad
D_7=-\frac{n_8}{2},\quad
D_8=\frac{n_7}{2},
\label{usualrel}
\end{equation}
and each component of ${\bf q}^{(16)}$ is identified with the number of
strings attached to each D7-brane.
After obtaining these relations one should deform the background
if one want to shift to the standard configuration.
Because ${\bf k}^{(16)}$, $m_\mu$ and $n_\mu$ are independent of the
Wilson lines ${\bf A}_\mu^{(16)}$,
once we have established the relations on the flat background,
we can obtain the relations between quantum numbers in heterotic string theory
with vanishing Wilson lines and invariant charges of junctions in standard configuration.
The ${\bf A}^{(16)}_\mu$ dependence (\ref{qandk}) of ${\bf q}^{(16)}$ corresponds to
the Hanany-Witten effect\cite{anomalous,BGL} and the invariant charges for
D-branes $Q_a$ in the standard configuration is
identified with the components of vector ${\bf k}^{(16)}$.
Furthermore we can represent the string charges $M_i$ and $N_i$ as follows\cite{JuncHet}.
\begin{equation}
M_0=k^1+\cdots+k^{16},
\end{equation}
\begin{equation}
M_1=-2m_8+2n_8-n_9-M_0,\quad
M_2=2m_8+2m_9-n_8-n_9+M_0,\quad
M_3=-2m_9-n_8+2n_9-M_0,
\label{m1m2m3}
\end{equation}
\begin{equation}
N_1=n_9,\quad
N_2=n_8-n_9,\quad
N_3=-n_8.
\label{n1n2n3}
\end{equation}
By using these relations, we can show that (\ref{bps}) is
equivalent with the BPS condition for junctions given in \cite{const,Geo}.


Next, let us discuss the compactification of heterotic string theory
on ${\bf T}^2_\ast$
and observe how the condition (\ref{bps}) is modified.
As we have explained in Section \ref{sec:Dualities},
the nontrivial structure of the gauge bundle
is realized by the Wilson line (\ref{sowl}).
However, for our purpose, it is convenient to use another subgroup
$Sp(1)\times Sp(8)$ and the following Wilson lines:
\begin{equation}
\mbox{$X^7$ direction} : i\sigma_z\otimes\exp(2\pi iA^a_7H_a),\quad
\mbox{$X^8$ direction} : i\sigma_x\otimes\exp(2\pi iA^a_8H_a),
\label{spwl}
\end{equation}
where $i\sigma_z$ and $i\sigma_x$ are elements of the $Sp(1)$ and
$H_a$ ($a=1,\ldots,8$) are the Cartan generators of the $Sp(8)$.
In this formulation, the gauge symmetry $Sp(8)$ appears when ${\bf A}^{(8)}_\mu=0$.
In order to obtain the modified BPS condition,
we have to quantize the worldsheet fermions taking account of the Wilson lines
(\ref{spwl}).
When both the winding numbers $n_8$ and $n_9$ are even,
the boundary condition of fermions is same as that of strings on ${\bf T}^2$
with Wilson line ${\bf A}^{(16)}_\mu=({\bf A}^{(8)}_\mu,{\bf A}^{(8)}_\mu)$
where ${\bf A}^{(8)}_\mu$ are Wilson lines of the $Sp(8)$ subgroup appearing in
(\ref{spwl}) and we can use the result for strings on ${\bf T}^2$.
The eight components of ${\bf A}_\mu^{(8)}$ represent the positions of
eight D7-branes on ${\bf T}^2/{\bf Z}_2^\ast$,
and the invariant charges representing the numbers of strings
attached to these D7-branes are charges
coupling to the ${\bf A}_\mu^{(8)}$.
The charge vector and normal ordered one are given by
\begin{equation}
{\bf q}^{(8)}={\bf q}^{(1)}+{\bf q}^{(2)},\quad
{\bf k}^{(8)}={\bf k}^{(1)}+{\bf k}^{(2)},
\label{qisqq}
\end{equation}
where ${\bf q}^{(i)}$ and ${\bf k}^{(i)}$ are former and latter half
of the charge vectors defined by (\ref{pdef}).
\begin{equation}
{\bf q}^{(16)}=({\bf q}^{(1)},{\bf q}^{(2)}),\quad
{\bf k}^{(16)}=({\bf k}^{(1)},{\bf k}^{(2)}).
\label{qqqkkk}
\end{equation}
Because ${\bf k}^{(16)}$ is on the self-dual even lattice,
${\bf k}^{(8)}$ takes values on the $Sp(8)$ root lattice generated by
\begin{equation}
-2{\bf e}_1,\quad
{\bf e}_1-{\bf e}_2,\quad
{\bf e}_2-{\bf e}_3,\quad
\ldots\quad,\quad
{\bf e}_7-{\bf e}_8.
\end{equation}
A state of a string on the torus ${\bf T}^2_\ast$ is specified by
giving winding numbers $n_\mu$, Kaluza-Klein indices $m_\mu$ and
$U(1)$ charges ${\bf q}^{(8)}$.
In the case of ${\bf T}^2$ compactification,
Kaluza-Klein indices $m_\mu$ are integral numbers.
However, now, we should take account of the effect of $Sp(1)$ Wilson
lines $i\sigma_z$ and $i\sigma_x$.
Because the eigen values for these operators on GSO even states are $\pm1$,
their effect is expressed by allowing $m_\mu$ to be half integers.
Let us define integral labels $\ol m_\mu=2m_\mu$.
Because the size of the ${\bf T}^2/{\bf Z}_2^*$
is half of the ${\bf T}^2/{\bf Z}_2$\cite{WithoutVector},
in order to normalize the coordinate on the ${\bf T}^2/{\bf Z}_2^\ast$
in the same way as the ${\bf T}^2/{\bf Z}_2$,
we introduce the vector $\ol{\bf A}^{(8)}_\mu=2{\bf A}^{(8)}_\mu$.
From (\ref{qandk}), the following relation between
${\bf q}^{(8)}$ and ${\bf k}^{(8)}$ is obtained.
\begin{equation}
{\bf q}^{(8)}
={\bf q}^{(1)}+{\bf q}^{(2)}
={\bf k}^{(1)}+{\bf k}^{(2)}-2{\bf A}_\mu^{(8)}n_\mu
={\bf k}^{(8)}-\ol{\bf A}_\mu^{(8)}n_\mu.
\label{q8k8}
\end{equation}
Using these variables, the quantization of Kaluza-Klein momenta (\ref{mpis}) is
rewritten as
\begin{equation}
m_\mu'=\frac{1}{2}\left(\ol m_\mu+{\bf k}^{(8)}\cdot\ol{\bf A}^{(8)}_\mu
                  -\frac{1}{2}\ol{\bf A}^{(8)}_\mu\cdot\ol{\bf A}^{(8)}_\nu n^\nu\right),
\label{mprimeis}
\end{equation}
where the inner products are defined with the metric ${\bf1}_8$.
By substituting these equation into (\ref{bpscond}),
we obtain the BPS condition for heterotic strings on the ${\bf T}^2_\ast$ as
\begin{equation}
2\ol m_7n_7+2\ol m_8n_8+{\bf k}^{(8)}\cdot{\bf k}^{(8)}+({\bf k}^{(1)}-{\bf k}^{(2)})^2\leq4.
\label{bps/20}
\end{equation}
Let us define a function $f({\bf k}^{(8)})$ as follows.
\begin{equation}
f({\bf k}^{(8)})=\min_{{\bf k}^{(8)}{\rm fixed}}(4-({\bf k}^{(1)}-{\bf k}^{(2)})^2).
\end{equation}
Because the vector ${\bf k}^{(16)}=({\bf k}^{(1)},{\bf k}^{(2)})$ is
on the self-dual even lattice,
we can represent $f({\bf k}^{(8)})$
in the following way.
\begin{equation}
f({\bf k}^{(8)})=|4-\#(\mbox{odd components of ${\bf k}^{(8)}$})|.
\label{fkdef}
\end{equation}
Because the sum of all the components of ${\bf k}^{(8)}$ is
even, $f({\bf k}^{(8)})$ is always $0$, $2$ or $4$.
Using this function, (\ref{bps/20}) is rewritten as
\begin{equation}
2\ol m_7n_7+2\ol m_8n_8+{\bf k}^{(8)}\cdot{\bf k}^{(8)}\leq f({\bf k}^{(8)}).
\label{bps/2}
\end{equation}
This is the modified BPS condition for heterotic strings on ${\bf T}^2_\ast$.
Until now, in the deduction of (\ref{bps/2}),
we have supposed $n_\mu\in2{\bf Z}$.
However, we can show that the same inequality is obtained for any
winding number as follows:
In the sector with $(n_7,n_8)=(1,0)\mod2$, the boundary condition
of worldsheet fermions is identical to that for
heterotic strings on ${\bf T}^2$ with Wilson lines
${\bf A}^{(16)}_8=({\bf A}^{(8)}_8+{\bf s}^{(8)}/2,{\bf A}^{(8)}_8-{\bf
s}^{(8)}/2)$
with ${\bf s}^{(8)}$ a eight-dimensional vector with all the
components $1/2$.
Because the BPS condition is independent of the Wilson lines, we have the
same result for this sector.
Furthermore, because other two sectors $(n_8,n_9)=(0,1)$ and
$(1,1)\mod2$ can be obtained from the $(1,0)$ sector by the modular
transformation of the torus,
the condition for these sectors should also be the same.

For heterotic strings on ${\bf T}^2$,
if a set of the quantum numbers $\{m_\mu,n_\mu,{\bf k}^{(16)}\}$ is specified,
there exists at least one state having the given quantum numbers,
and if the set of quantum numbers satisfies the BPS condition,
at least one BPS state exists.
However, for heterotic strings on ${\bf T}^2_\ast$,
it is not the case.
For example, let us consider states with $n_7=n_8=0$
and ${\bf k}^{(8)}\cdot{\bf k}^{(8)}=4$.
For these states to be BPS states, all the components of ${\bf k}^{(8)}$
should be even or odd,
and ${\bf k}^{(8)}$ should have one nonzero component $\pm2$.
There are $16$ such states $\lambda_{-1/2}^{a\pm}\lambda_{-1/2}^{a+8\pm}|0\rangle$.
These are what become massless when $Sp(8)$ gauge group is enhanced
on the $O^+$
and correspond to the long roots of $Sp(8)$ Lie algebra.
(Generally, $Sp(k)$ has $2k$ long roots with norm $4$.)
Because the eigenvalues of $i\sigma_x$, $i\sigma_z\in Sp(1)$ for these
states are $+1$, the quantum numbers $\ol m_\mu$ should be even
integers. 
In other words, although the set of quantum numbers $\{\ol m_\mu,n_\mu,{\bf k}^{(8)}\}$
with $n_\mu=0$, ${\bf k}^{(8)}\cdot{\bf k}^{(8)}=4$ and $\ol m_8\ol m_9\notin{\bf4Z}$
satisfies the BPS condition (\ref{bps/2}),
there are no BPS states with the quantum numbers.
(As a result, we have $Sp(8)$ gauge group only on
one orientifold plane.)

\section{Relation between quantum numbers}\label{sec:1to1}

In this section, we will explain how we can obtain
the one to one correspondence between quantum
numbers of
heterotic strings on ${\bf T}^2_\ast$ and invariant charges
of junctions on ${\bf T}^2/{\bf Z}_2^\ast$.
It is an variation of the relation between Het$/{\bf T}^2$ and
IIB$/({\bf T}^2/{\bf Z}_2)$ given in \cite{JuncHet}.

Via the duality explained in Section \ref{sec:Dualities},
the D-string winding numbers $D_\mu$ and the F-string winding numbers
$F_\mu$ on the flat ${\bf T}^2/{\bf Z}_2^\ast$ is related
to the quantum numbers of heterotic strings on ${\bf T}^2_\ast$ as follows:
\begin{equation}
F_7=2m_7',\quad
F_8=2m_8',\quad
D_7=-\frac{1}{2}n_8,\quad
D_8=\frac{1}{2}n_7.
\label{unwindcorr}
\end{equation}
Comparing to (\ref{usualrel}),the reason for the existence of extra
$2$ factors for $F_\mu$ is that,
because the size of the ${\bf T}^2/{\bf Z}_2^\ast$ is half of that of
${\bf T}^2/{\bf Z}_2$,
in order to give the same energy the winding numbers should be twice
as (\ref{usualrel}).
These extra $2$ factors are compensated by the extra $1/2$ factor in
(\ref{mprimeis}).
Therefore, it is expected that we obtain the exactly same relations between
$\{\ol m_\mu,n_\mu,{\bf k}^{(8)}\}$ and $\{D_\mu,F_\mu\}$ as that
for the Het$/{\bf T}^2$ - IIB$/({\bf T}^2/{\bf Z}_2)$ duality given in \cite{JuncHet}.
However, there is one problem.
To use the relation (\ref{unwindcorr}),
we should use the flat background.
Because, unlike the orientifold with four $O^-$,
the positive R-R charge of the $O^+$ cannot be canceled by the D7-brane
charge, there is no BPS flat background.

This obstruction can be overcome by using a non-BPS flat background as follows:
If we allow non-BPS configurations,
we can make a flat background
by setting four anti-D7-branes on top of the $O^+$.
\begin{figure}[htb]
\begin{displaymath}
\put(90,-10){$X^7$}
\put(-15,100){$X^8$}
\put(30,65){$(F_7,D_7)$}
\put(35,120){$(F_8,D_8)$}
\put(20,30){$\ol{\rm D7}\times4$}
\put(20,145){${\rm D7}\times4$}
\put(125,30){${\rm D7}\times4$}
\put(125,145){${\rm D7}\times4$}
\put(0,0){$O^+$}
\put(0,170){$O^-$}
\put(170,0){$O^-$}
\put(170,170){$O^-$}
\epsfbox{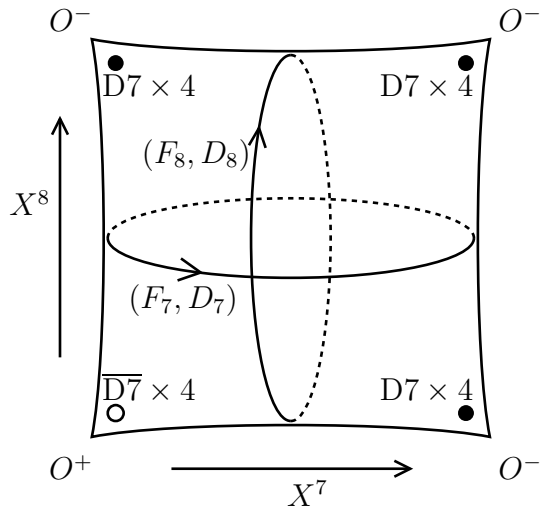}
\end{displaymath}
\caption{The flat background with three $O^-$ and one $O^+$.
Each bullet represent a stack of four D7-branes and a circle is a set of four anti-D7-branes.}
\label{d7d8.eps}
\end{figure}
This brane configuration gives the gauge group $SO(8)^4$ with rank $16$.
This implies that vectors ${\bf k}^{(8)}$, ${\bf q}^{(8)}$ and $\ol{\bf A}^{(8)}_\mu$
should be extended to $16$-dimensional vectors.
This can be done in the following way.
Let us consider the situation in which $d$ D7-branes and $d$ anti-D7-branes are created
at a certain point on the ${\bf T}^2/{\bf Z}_2^\ast$.
Then $8$ dimensional vectors should be extended to $8+2d$ dimensional vectors.
Let us define ${\bf A}^{(8+d,d)}_\mu$ as vectors representing the positions
of branes and anti-branes.
We assume the former $8+d$ components of ${\bf A}^{(8+d,d)}$ represent
the positions of $8+d$ D-branes
and latter $d$ components represent the positions of the $d$ anti-D-branes.
If the $d$ D-branes and the $d$ anti-D-branes are stacked at a point,
and junction does not attached on them,
the quantization (\ref{mprimeis}) should not change.
This is achieved by introducing the metric
$g_{ab}=\diag({\bf1}_{8+d},-{\bf1}_d)$.
Furthermore, if we demand that
when a D-brane and an anti-D-brane are on top of each other,
an open string connecting them should not contribute to the
string winding number obtained by (\ref{mprimeis}),
we should define ${\bf q}^{(8+d,d)}$ as a vector
whose former $8+d$ components represent the number of strings
going out from each D-brane while latter $d$ components
represents the number of strings
going into each anti-D7-brane.
Under this definition, $\ol{\bf A}_\mu^{(8+d,d)}$ dependence
of ${\bf q}^{(8+d,d)}$
representing the Hanany-Witten effect can be given in the same equation
as (\ref{q8k8}):
\begin{equation}
{\bf q}^{(8+d,d)}={\bf k}^{(8+d,d)}-\ol{\bf A}^{(8+d,d)}_\mu n_\mu.
\end{equation}

In this way, we can extend the vectors to $16$-dimensional vectors ($d=4$)
and we can realize the flat background, on which
the relations (\ref{unwindcorr}) are available.
Once we have obtained the relations (\ref{unwindcorr}),
in the same way as we explained in Section \ref{sec:BPS}
in the case of Het$/{\bf T}^2$ and IIB$/({\bf T}^2/{\bf Z}_2)$,
we can obtain the following relation for the string charges of junctions
on the background with $12$ D7-branes and $4$ anti-D7-branes
near the $O^+$.
\begin{equation}
M_0=\sum_{a=1}^{12}k'^a-\sum_{a=13}^{16}k'^a.
\label{m0is}
\end{equation}
\begin{equation}
M_1=-2\ol m_8+2n_8-n_9-M_0,\quad
M_2=2\ol m_8+2\ol m_9-n_8-n_9+M_0,\quad
M_3=-2\ol m_9-n_8+2n_9-M_0,
\label{ms}
\end{equation}
\begin{equation}
N_1=n_9,\quad
N_2=n_8-n_9,\quad
N_3=-n_8.
\label{ns}
\end{equation}
(We use $k'^a$ to represent the components of the ${\bf k}^{(8+d,d)}$,
while the components of ${\bf k}^{(8)}$ are $k^a$.)

To obtain the standard configuration with eight D7-branes,
four pairs of D7-branes and anti-D7-branes should be annihilated.
In this process, the gauge group $U(12)\times U(4)$ is broken to $U(8)$
by means of the tachyon condensation\cite{tachyon}.
After the condensation, we have eight charges $k^a$ instead of $k'^a$.
Unless we specify the pattern of the tachyon condensation,
we cannot represent $k^a$ by $k'^a$.
However,
because the $U(1)$ is not broken by the tachyon condensation,
the total charge coupled with the diagonal $U(1)$ is unchanged
and the equation (\ref{m0is}) is
replaced by
\begin{equation}
M_0=\sum_{a=1}^8k^a.
\label{m0}
\end{equation}
Because the relations (\ref{m0}), (\ref{ms}) and (\ref{ns}) are
identical to the relation appearing
in Het$/{\bf T}^2$-IIB$/({\bf T}^2/{\bf Z}_2)$ duality,
the left hand side of (\ref{bps/2}) must be
rewritten to the identical expression to what is given in \cite{JuncHet}.
As a result, we obtain the following BPS condition for junctions:
\begin{equation}
({\bf J},{\bf J})\leq f({\bf k}^{(8)}),
\end{equation}
where the left hand side is the self intersection number of junction ${\bf J}$
obtained by means of the formula given in \cite{arbitrary}
and $f({\bf k}^{(8)})$ is the function defined by (\ref{fkdef}).
However, as we mentioned above, we should note that even this equation hold,
the existence of the corresponding BPS junction is not guaranteed.

Let us consider the decompactification limit in the type IIB theory
and look at the vicinity of $O^+$.
We suppose $r$ of eight D7-branes are left near the $O^+$ and other $8-r$
D7-branes are decoupled.
In this limit, only the states with $\ol m_\mu=n_\mu=k^{r+1}=k^{r+2}=\cdots=k^8=0$
are left.
The BPS condition is given by
\begin{equation}
({\bf J},{\bf J})={\bf k}\cdot{\bf k}\leq f({\bf k}),
\end{equation}
where ${\bf k}$ is $r$-dimensional charge vector on the $Sp(r)$ root
lattice representing
the numbers of the strings attached to the $r$ D7-branes.
Although the function $f({\bf k})$ was originally defined for the eight
dimensional vector, we can use the same definition for the vector of
dimension $r<8$
because the definition (\ref{fkdef})
is unaffected by extra zero components.
Although we cannot use the heterotic / type IIB duality
for $r>8$, let us assume the same BPS condition holds.
Then we find that the adjoint representation of $Sp(r)$
gauge group are reproduced as follows:
Let $n_{\rm odd}$ denote the number of odd components of a vector ${\bf k}$.
The norm ${\bf k}\cdot{\bf k}$ is always equal with or lager than $n_{\rm odd}$
and the following inequality should hold for BPS states.
\begin{equation}
n_{\rm odd}\leq{\bf k}\cdot{\bf k}\leq|4-n_{\rm odd}|.
\end{equation}
Due to this equation, only two values $0$ and $2$ is allowed for $n_{\rm odd}$.
When $n_{\rm odd}=0$, the vector ${\bf k}$ should satisfy ${\bf k}\cdot{\bf k}\leq4$.
Therefore, all the components are zero or only one component is $\pm2$ and others are zero.
They give the Cartan generators and the long roots of $Sp(r)$ gauge group.
When $n_{\rm odd}=2$, two components of ${\bf k}$ are $\pm1$ and
they correspond to the short roots of $Sp(r)$.
Collecting these states altogether, we obtain the adjoint
representation
of the $Sp(r)$.

\section{Conclusions and discussions}
We investigated the duality between heterotic strings
on torus with nontrivial gauge bundle structure ${\bf T}^2_\ast$
and junctions on background containing an $O^+$.
First, we gave the BPS condition for heterotic strings on ${\bf T}^2_\ast$,
Using the correspondence between quantum numbers of heterotic strings
and junctions established on a non-BPS flat background,
we obtained the modified BPS conditions for junctions on ${\bf T}^2/{\bf Z}_2^\ast$.
By means of this condition, we could reproduce the adjoint representation of
$Sp(r)$ gauge group on the $O^+$.

As we mentioned in Introduction,
string junctions are very useful tools for analyzing non-perturbative
spectra of supersymmetric Yang-Mills theories on the probe D3-branes.
The gauge theory on $n$ D3-branes on a background with an $O^+$ and $k$ D7-branes
is ${\cal N}=2$ supersymmetric $SO(2n)$ gauge theory with $k$ hypermultiplets.
For complete analysis of this theory,
we need the BPS conditions for junctions
attached on the probe D3-branes on a background with an $O^+$.
If we had obtained it, we would be able to know the complete spectrum of BPS particles
of the theory.
However, unfortunately,
our argument with the perturbative heterotic strings
is not sufficient for the purpose
because dual of D3-branes are solitonic five-branes and
our argument here did not contain them.

\section*{Acknowlidgements}

I would like to thank S. Sugimoto and I. Kishimoto for helpful
conversations.
This work is supported in part by a Grant-in-Aid for Scientific
Research from the Ministry of Education, Science, Sports and Culture
(\#9110).



\begin{thebibliography}{10}
\bibitem{fromopen}
    M.~R.~Gaberdiel and B.~Zwiebach,
    Nucl.Phys.{\bf B518}(1998)151,
    {\tt hep-th/9709013},\\
    {\em``Exceptional groups from open strings.''}
\bibitem{arbitrary}
    O.~DeWolfe, B.~Zwiebach,
    Nucl.Phys.{\bf B541}(1999)509,
    {\tt hep-th/9804210},\\
    {\em``String Junctions for Arbitrary Lie Algebra Representations.''}
\bibitem{affine}
    O.~DeWolfe,
    {\tt hep-th/9809026},\\
    {\em``Affine Lie Algebras, String Junctions and 7-Branes.''}
\bibitem{uncovering}
    O.~DeWolfe, T.~Hauer, A.~Iqbal and B.~Zwiebach,
    {\tt hep-th/9812028},\\
    {\em``Uncovering the Symmetries on [p,q] 7-branes:
    Beyond the Kodaira Classification.''}
\bibitem{km}
    O.~DeWolfe, T.~Hauer, A.~Iqbal and B.~Zwiebach,
    {\tt hep-th/9812209},\\
    {\em``Uncovering Infinite Symmetries on [p,q] 7-branes:
    Kac-Moody Algebras and Beyond.''}
\bibitem{flavor}
    Y.~Imamura,
    Phys.Rev.{\bf D58}(1998)106005,
    {\tt hep-th/9802189},\\
    {\em``$E_8$ flavor multiplets.''}
\bibitem{const}
    O.~DeWolfe, T.~Hauer, A.~Iqbal and B.~Zwiebach,\\
    Nucl.Phys.{\bf B534}(1998)261, {\tt hep-th/9805220},\\
    {\em``Constraints on the BPS Spectrum of N = 2, D = 4 Theories with A-D-E Flavor Symmetry.''}
\bibitem{Geo}
    A.~Mikhailov, N.~Nekrasov and S.~Sethi,
    Nucl.Phys.{\bf B531}(1998)345,
    {\tt hep-th/9803142},\\
    {\em``Geometric Realizations of BPS States in N=2 Theories.''}
\bibitem{su2anomaly}
    E.~Witten,
    Phys.Lett.{\bf117B}(1982)324,\\
    {\em``An SU(2) Anomaly''}
\bibitem{JuncHet}
    Y.~Imamura,
    Prog.Theor.Phys. {\bf101} (1999) 1155,
    {\tt hep-th/9901001},\\
    {\em``String Junctions and Their Duals in Heterotic String Theory.''}
\bibitem{WithoutVector}
    E.~Witten,
    JHEP {\bf9802} (1998) 006,
    {\tt hep-th/9712028},\\
    {\em``Toroidal Compactification Without Vector Structure.''}
\bibitem{BGL}
    O.~Bergman, M.~R.~Gaberdiel and G.~Lifschytz,\\
    Nucl.Phys. {\bf B524} (1998) 524,
    {\tt hep-th/9711098},\\
    {\em``String Creation and Heterotic-Type I' Duality.''}
\bibitem{anomalous}
    C.~P.~Bachas, M.~R.~Douglas and M.~B.~Green,
    JHEP {\bf9707} (1997) 002,
    {\tt hep-th/9705074},\\
    {\em``Anomalous Creation of Branes.''}
\bibitem{tachyon}
    A.~Sen,
    JHEP {\bf9808} (1998) 012,
    {\tt hep-th/9805170},\\
    {\em``Tachyon Condensation on the Brane Antibrane System.''}
\end{thebibliography}
\end{document}